\documentclass[11pt,a4]{article}
\usepackage{graphicx,amsmath}

\def\more{\raisebox{-1.1mm}{$\stackrel{>}{\sim}$}}

\begin{document}

\title{Intermediate Mass Stars $\leftrightarrow$ Massive Stars. \\
\Large
A workshop around causes and consequences of differing evolutionary paths}
\author{E. Josselin$^1$, A. Lan\c con$^2$ \\
\footnotesize
$^1$ GRAAL, Universit\'e Montpellier II \& CNRS, 34095 Montpellier, France \\
\footnotesize 
$^2$ Observatoire Astronomique de Strasbourg, Universit\'e Louis Pasteur \& CNRS, Strasbourg, France 
}
\date{Strasbourg (France), February 9-11, 2009}
\maketitle

\section*{Foreword}
The post-main sequence evolution of stars of intermediate or large masses is notoriously complex. 
In the recent past, a number of workshops and meetings have focused on either the Asymptotic 
Giant Branch of intermediate mass stars, or the evolution of massive stars. But how well defined is the 
boundary between these categories of objects defined? How would an observer proceed to 
classify stars into one or the other category? How do objects near the boundary evolve, die, and 
contribute to the chemical evolution of their environment?

During this 3-day international workshop\footnote{Supported by the PNPS (Programme National de 
Physique Stellaire, CNRS, France) and ANR (Agence Nationale de la Recherche, France). }, 
26 high quality presentations were
given by specialists in the relevant fields of astrophysics, and stimulating
discussions followed. It is technically impossible to provide an exhaustive
census of the results and ideas that emerged. In this brief article,
we choose to point to key elements of the workshop, some of which are
now the topic of new collaborations and will lead to publications elsewhere.
For the sake of brevity, we deliberately cite only the contributors to the 
workshop and no external references. Many bibliographic references can be found 
in the original presentations, which can be retrieved through:
http://astro.u-strasbg.fr/observatoire/obs/stars2009/stars2009.html 
The programme workshop, which includes the titles of the individual 
contributions, is provided as an appendix.

\begin{figure}[h]
\begin{center}
\includegraphics[width=7cm,bbllx=0,bblly=150,bburx=594,bbury=701,clip=]{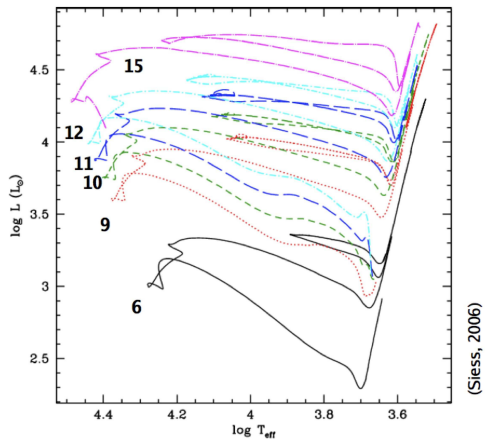}
\caption{Evolutionary tracks from the STAREVOL code (from \it{Siess \& Palacios}).}
\label{siess3}
\end{center}
\end{figure}

\section{Definition, mass and impact of super-AGB stars}
\subsection{The mass domain}
\begin{figure}[h]
\begin{center}
\includegraphics[angle=90,width=8cm]{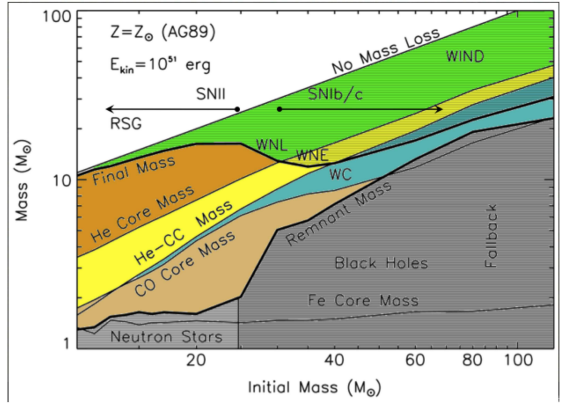}
\caption{Fate of massive stars (from {\it Limongi}). }
\label{limongi}
\end{center}
\end{figure}

``Low-mass'' stars end as CO white dwarfs, ``massive" stars, which are at the 
origin of core-collapse type II supernovae (SNe), go through all the nuclear burnings, from 
H to Si (see Fig. \ref{limongi} for the fate of massive stars, and {\it Limongi} for a detailed 
discussion). In between, one meets a domain of stellar masses, which evolve through the 
{\it super asymptotic giant branch} (super-AGB) phase and will end as either ONe white dwarfs 
or electron capture SNe. 

\begin{figure}[h]
\begin{center}
\includegraphics[angle=90,width=8cm]{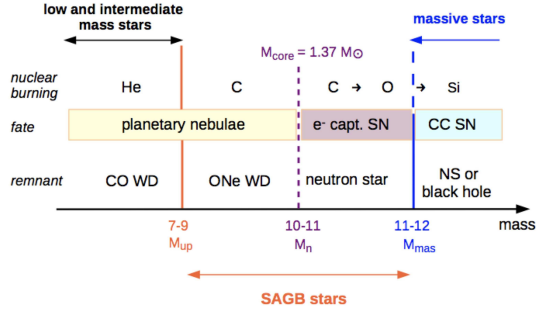}
\caption{Definition of the threshold masses (from {\it Siess \& Palacios}). }
\label{siess1}
\end{center}
\end{figure}

What defines super-AGB stars is what happens in the core. Three critical masses thus have to be 
determined: the minimum initial stellar mass (referred hereafter as $M_{up}$) which will allow 
carbon ignition in the core, the minimum mass $M_n$ for neutron star formation and 
the limit $M_{mas}$ for type II SN (we adopt here the notation proposed by {\it Siess \& Palacios}; 
see Fig. \ref{siess1}). 

The value of the {\it core} mass corresponding these different transition masses seems rather well 
determined by theory: 1.05 $M_odot$ is the minimum core mass for carbon ignition, above 1.37 
$M_odot$ the (degenerate) neon-oxygen core is massive enough to activate electron capture (EC) 
reactions leading to the formation of a neutron star and if the core mass at the end of He-burning 
exceeds the Chandrasekhar mass (1.44 $M_odot$), then the star will proceed through all nuclear 
burning stage and end up as a core collapse SN (SNII or SNIbc depending on the initial mass and 
mass loss). But going back to initial masses is more problematic. 
Treatment of (semi-)convection (and associated overshooting and dredge-up), mass loss and 
carbon burning are among the most influential ingredients. 
While the differences between critical masses are reasonably robust against the input 
physics (e.g. adopted nuclear rates) and models, at least for {\it single}  star evolution 
($M_{mas} - M_{up} \, \approx 2$~M$_\odot$, $M_{mas} - M_n \, \approx 0.25$~M$_\odot$ 
at solar metallicity; {\it Weiss, Pols}), their absolute values are quite controversial. 

{\it Straniero} presented a new study of the 
$^{12}$C($^{12}$C$,p)^{23}$Na rate (with the discovery of two new resonances) which reduces 
$M_{up}$ by 2 M$_\odot$, from 8 to 6 M$_\odot$. Taking into account overshooting has a a 
similar impact ({\it Siess \& Palacios}). And the uncertainty on the evolution of massive stars is still 
largely dominated by the poor kowledge of the $^{12}$C$(\alpha,\gamma)^{16}$O cross section 
({\it Limongi}).

\subsection{Constraints from white dwarfs and supernovae}
Observational constraints on the limiting masses are thus required. In principle, they can be 
obtained from the observation of white dwarfs (WD) in open clusters ({\it Meynet, Kalirai, Williams}). 
This method goes through several steps: based on atmosphere models, determination of WD 
effective temperature and gravity; from a mass-radius relation, estimation of WD masses; 
from  theoretical isochrones, estimation of the age of the cluster. Since the age of the cluster is also 
the sum of nuclear lifetime of the progenitor of the white dwarf and its cooling age, one deduces 
the nuclear lifetime of the progenitor, and thus its initial mass.  Obviously, all these steps require 
theoretical inputs ... so the results are model dependent! Given the Main Sequence lifetime of a 
8 M$_\odot$ ($\sim$ 40 Myrs), 
this method requires the study of open clusters with an age of a few 10 Myrs. 
{\it Williams} obtained a {\it lower limit} on $M_{mas}$ of 6.3-7.1 M$_\odot$. 

The knowledge of the chemical composition of white dwarfs is crucial for determining the cooling 
rate and the progenitor mass. Strong mass loss on the red giant branch may lead to a bypass of the 
AGB, and thus the formation of a He white dwarf, which will cool 3 times more slowly than a
CO white dwarf. This could explain the apparent discrepancy between Main Sequence turnoff age 
and white dwarf cooling age for some clusters ({\it Kalirai}). Hot DQ white dwarfs may be the progeny 
of 9-11 M$_\odot$ stars, and thus have a ONe core. This could be tested with asteroseismology 
({\it Williams}). 

The study of type II-P SNe and the identification of their progenitors 
(mostly red supergiant stars) should constrain $M_{mas}$ ({\it Smartt, Eldridge}). A lower limit 
of 8.5 (+1/-1.5) M$_\odot$ is found. Recent observations of
possible electron-capture supernovae such as supernova 2008S and the WD
masses of Kurtis et al. {\it Eldridge} estimated that $M_{mas} - M_n = 1.1 M_\odot$. 

Combining these theoretical and ``observational" results may lead to inconsistencies. Apart 
from improvements in models and interpretation of observations, one should consider the possible 
effects of binarity. Indeed, as emphasised by {\it Pols}, merging in close binaries leads to a decrease  
of effective mass limit. 

\begin{figure}[h]
\begin{center}
\includegraphics[angle=90,width=8cm]{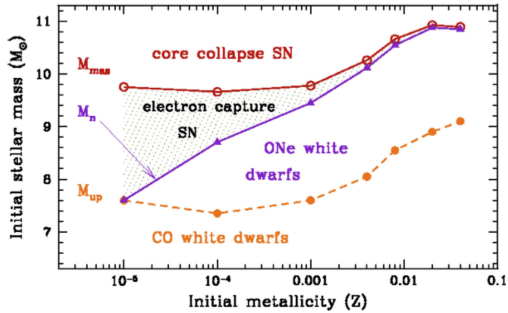}
\caption{Evolution of threshold masses as a function of metallicity (from {Siess  \& Palacios}). }
\label{siess2}
\end{center}
\end{figure}

Finally, one should keep in mind that these threshold masses depend on metallicity, as illustrated 
in Fig. \ref{siess2} (extracted from {\it Siess \& Palacios}). 

\subsection{Stellar populations and yields}
How important are super-AGB stars for stellar populations and Galactic chemical evolution? 

Most evolutionary tracks available for population synthesis purposes are still incomplete at
the masses relevant to super-AGB stars, lacking the latest, reddest and also brightest phases of evolution. This may lead to unrealistically blue colours for synthetic populations at ages around 
100 Myr. 

Despite the relatively narrow mass interval ($\Delta M = 2$~M$_\odot$), a Salpeter-like Initial 
Mass Function gives a  number of stars with $M_{up} < M < M_{mas}$ which is about 2/3 
that of the stars with $M > M_{mas}$ (the progenitors of core collapse SNe), and these may thus play 
an important role in chemical evolution. 

Super-AGB stars are the most massive stars which experience the second dredge-up 
({\it Siess \& Palacios}), but maybe no third dredge-up. Hot Bottom Burning is thus a   
key-ingredient in determining the super-AGB yields and these stars may be strong contributors 
of early  $^{14}$N and$^{13}$C enrichment (up to Z = $10^{-4}$; {\it Chiappini}). 
Such yields may explain the chemically anomalous  second generation of stars in globular 
clusters ({\it d'Antona}), although this remains a matter of debate, and more massive 
rotating stars are an alternative favoured by some ({\it Decressin}). 

Concerning elements heavier than Fe, the timescale and importance of neutron exposure 
in super-AGB stars are small, and the main neutron source is $^{22}$Ne($\alpha$,n)$^{25}$Mg. 
This leads to significant production of  a limited number of elements, around Z=40, such as Rb 
({\it Straniero, Frishknecht}). 

Many open questions remain. In particular, what are the final C/O ratio in super-AGB stars. 
Is it possible to find carbon-rich super-AGB stars ? If yes, at which metallicities ? 

\section{Identification of super-AGB stars: atmospheric properties}
\subsection{Atmospheres and spectral classification}
If one excepts asteroseismology, still applicable to a limited number of stars, only the atmosphere 
of stars are observable, so diagnostics on their nature rely on our ability to interpret their 
spectra. The most widely used models (e.g. MARCS or PHOENIX) are based on ``classical" 
approximations: 1D, static atmospheres, LTE, mixing length theory for convection, a 
depth-independant microturbulent velocity. From such models, it is noticeable that 
red supergiant stars and AGB stars with same luminosity and temperature, thus differing 
only by their masses (and thus their surface gravity), have a different thermal structure, and 
different emerging spectra 
({\it Plez}). However, pulsations and/or convection, probed by photometric and line profile variability, 
make the validity of the classical approximations 
questionable. 3D models are under development, but still need improvements (temperature 
profiles are too shallow, velocity fields are too small; {\it Plez}). In parallel, progress in 
full NLTE 3D radiative transfer aims at producing more reliable diagnostics ({\it Hauschildt}). 

Meanwhile, the use of classical models encountered success, such as the new 
determination of effective temperatures of red supergiant stars for different metallicities 
({\it Levesque}). Remarkably, this new calibration agrees well with independent 
determinations,  based e.g. on interferometric measurements ({\it Gray}). It also shows a better 
agreement with evolutionary tracks (Geneva model) ... which may be partly fortuitous. 
(The end points of the STARS models extend to cooler temperatures
than the Geneva models and also the observed positions of Galactic
RSGs. However it is important to compare models which end at the
same burning stage to ensure reliable physical comparisons. Also models
spend more time at the bottom of their RSG track and accelerate up it,
spending more time in the region of the HR diagram inhabited by observed
RSGs; {\it Smartt \& Eldridge}). 
Let us recall that the effective temperature at the red supergiant stage depends on the 
ratio of the mixing length to the pressure scale height, which is a free parameter 
in general adjusted using solar models.
Combination of realistic stellar atmospheres and evolutionary models, to obtain 
reliable predictions of effective temperatures, are needed, but this is far from being an easy task... 
Another improvement of both atmospheric and evolutionary models would concern the 
treatment of convection, based in both cases on the mixing length theory (MLT). The used 
$\alpha$ parameter is widely that calibrated with the Sun. To what extent is it applicable 
to red (super-)giant stars? Is MLT itself appropriate? 

The spectral classification of M-type stars relies on the {\it absolute} strength 
of TiO bands, and are thus metallicity dependent ({\it Gray}). To overcome this problem 
and obtain a  univocal correspondence between spectral type and effective temperature, 
a possibility could be to define a spectral classification based on atomic line or molecular 
band ratios. Is such a revision in spectral classification desirable? 

In the context of (super-)AGB stars, an extension of the calibration of effective temperatures 
toward lower mass/luminosity, longer wavelengths (infrared) and lower metallicities is highly 
desirable. Static models seem to reproduce satisfactorily the spectra of the warmer stars 
($T_{eff} \more 3500$~K) but, as one goes toward lower temperatures, one is confronted 
with inconsistencies between the visible and the infrared parts of the spectra ({\it Lan\c con}). 
Finally, the link between the empirical concept of luminosity class and stellar mass or 
evolutionary models is pending. 

As mentioned above, another  approach to determine fundamental stellar parameters consists 
in the interpretation of interferometric observations. Realistic (1D) hydrodynamic models now 
allow to interpret consistently such observations, without invoking {\it ad-hoc} structures such as 
``molspheres" ({\it Wittkowski, Sacuto}). They indeed show how extended the atmospheres of 
(super-)giant stars are. In return, this renders the concept of effective temperature less clear 
for observers, as it cannot be considered as a ``surface" temperature. 

\subsection{Pulsations, mass loss and circumstellar envelopes}
Apart from complicating the building of reliable stellar atmosphere models, stellar pulsations 
are interesting for themselves. In particular, the apparent universality (i.e. metallicity 
independent) of the period-luminosity relations is remarkable ({\it Schultheis}). There 
remains work to do. The origin of the variability of some classes of objects (in particular the 
sequence ``D") remain mysterious. A complete census of semi-regular variables in the solar 
neighbourhood is still missing, and would bring important constraints for stellar evolution models. 

Pulsation is a key-ingredient in the mass-loss process of AGB stars, together with radiation on 
dust. From Hydrodynamical Dust driven wind models, it appears that abundances are also 
important. As metallicity increases, similar mass-loss rate, larger wind velocity and 
dust-to-gas ratio are predicted for a given C/O value, while mass-loss rate, wind velocity and 
dust-to-gas ratio all increase with C/O ({\it Groenewegen}). High angular 
resolution observations also point toward the need of 3D models, as it appears that 
asymmetries in the wind of AGB stars develop very early ({\it Chesneau}). 

A good understanding of the mass-loss process, and a proper determination of the total 
mass-loss rate, is fundamental to predict initial-final mass relations. Intermediate-mass stars 
are supposed to experience stronger winds and have thus larger amounts of circumstellar 
material, compared to more massive stars. Then how many evolved intermediate mass stars 
are missed because of extinction? What place do OH/IR stars occupy in the context of intermediate 
mass star and super-AGB evolution? 

\section{The future}
Super-AGB stars represent a real challenge for stellar physics. More generally, 
intermediate mass stars, and super-AGB stars in particular, play an important role 
in ``the Bigger Picture" ({\it Kalirai}). The evolution in different environments, such as the correlation 
between mass loss and metallicity (i.e. how extinct extra-galactic super-AGB stars may be?), the 
metallicity of bulge stars, are key-ingredients for the interpretation of light from distant galaxies, or 
our understanding of the Milky Way formation. 

From an observer's point of view, upcoming facilities will help clarify some of the questions raised 
above. Important  projects mentioned during this workshop include: the LSST (Large Synoptic Survey 
Telescope, e.g. HR diagrams of clusters), ALMA (Atacama Large Millimetre Array, mass loss in the Magellanic Clouds as a  test dust driven wind theory), second generation of VLTI instrumentation 
(wind aymmetries), GAIA (survey of long period variables to constrain pulsation properties in the Milky 
Way better). 

We thus hope and believe that this workshop will be followed by other ones. Keep in touch!

\newpage

\appendix

\section*{Programme of the workshop}

{\bf Day 1 - Monday, February 9, 2009:} \\
Topics: \\
- AGB stars, super-AGB stars and related objects. Definitions. Which evolutionary path to each of these categories? \\
- Predicted properties from evolutionary models. Model assumptions and their effects. \\

\begin{itemize}
\item Achim Weiss - Review about location in the theoretical diagram, lifetimes, surface 
abundances and the like 
\item Oscar Straniero - Around Mup : energy generation, energy sink and nucleosynthesis 
\item Ana Palacios - Nucleosynthesis and the fate of super-AGB stars 
\item Georges Meynet - Mass transition between white dwarf progenitors and neutron 
stars progenitors - Present status and perspectives 
\item Onno Pols  - Effects of binaries on the transition mass and on the occurrence of 
electron-capture supernovae 
\item Steve Smartt - Observational constraints on the progenitors of type II-P supernovae 
\item J. Eldridge - Constraints from combining SN-progenitors, white dwarf masses \& stellar models 
\item Workshops and discussion 
\begin{enumerate}
	\item Comparisons between spectral models 
	\item Spectral typing of model spectra 
	\item Do we know the mass-cut between WD and NS progenitors? 
	\item Comparisons between evolutionary models 
\end{enumerate}
\item Marco Limongi - massive stars : pre-SN evolution, explosion, nucleosynthesis 
\item Jason Kalirai - Initial-final mass relations 
\item Kurtis A. Williams - Observations of white dwarf remnants of intermediate mass stars 
\end{itemize}

\bigskip 

{\bf Day 2 - Tuesday, February 10, 2009:} \\
Topics: \\
- Predicted and observed atmospheric properties. Comparisons with observations. \\

\begin{itemize} 
\item Bertrand Plez - Differences between atmospheres of intermediate mass stars and of 
massive stars, recent model developments 
\item  Richard Gray - Spectral types in the upper right of the HR diagram Ð known relations 
to physical properties, known problems; the future of spectral types 
\item Peter Hauschildt - Lessons from 3D simulations, views on micro/macroturbulence 
\item Workshops and discussion 
\item Emily Levesque - Modelling the physical properties of red supergiants 
\item Ariane Lan\c con - Near-IR spectra of luminous red stars. Data versus models. 
\item Eric Josselin - Water and other molecules. Observational evidence for molecular 
layers. 
\item Mathias Schultheis - Period-luminosity relations on the AGB 
\item Markus Wittkowski - Interferometric data compared to model atmospheres 
\item Olivier Chesneau - Constraints from interferometry in the mass range of interest to 
the workshop 
\item S\'ephane Sacuto - Hydrodynamic interpretation of AGB stars observed at high 
angular resolution. 
\end{itemize}

{\bf Day 3 - Wednesday, February 11, 2009:} \\
Topics: \\ 
- Chemistry, yields, contributions to the chemical evolution of galaxies \\
- Summary of the sub-workshops \\

\begin{itemize} 
\item Martin Groenewegen - Mass loss from AGB stars and red supergiants in the 
Magellanic Clouds. 
\item Cristina Chiappini - Chemical evolution of galaxies with and without AGB stars or 
super-AGB stars 
\item Thibaut Decressin - The role of super-AGB stars and related objects in the self-enrichment of star clusters. 
\item Francesca D'Antona - A dynamical model for the formation of multiple populations 
in globular clusters, and the role of super-AGBs. 
\end{itemize}

\end{document}